\documentclass[useAMS,usenatbib]{mn2e}
\usepackage{graphicx}
\title[NuSTAR observation of V745 Sco]{A NuSTAR observation of the fast symbiotic nova V745 Sco
in outburst}
\author[M. Orio, V. Rana, K. L. Page, J. L. Sokoloski, and F. Harrison ]
{M. Orio $^{1,2}$\thanks{E-mail:marina.orio@oapd.inaf.it},
V. Rana$^{3}$\thanks{}, K. L. Page$^{4}$\thanks{}, J. Sokoloski$^{5}$\thanks{}, F.
Harrison$^{3}$\thanks{}\\
 $^1$ INAF--Osservatorio di Padova, vicolo dell' Osservatorio 5,
   I-35122 Padova, Italy \\
 $^2$ Department of Astronomy, University of Wisconsin, 475 N. Charter Str., Madison WI 53704\\
 $^3$ Cahill Center for Astronomy and Astrophysics, California Institute of Technology, Pasadena, CA 91125, USA \\ 
 $^4$ Department of Physics and Astronomy, University of Leicester, Leicester, LE1 7RH, UK \\
 $^5$  Columbia Astrophysics Laboratory, Columbia University, 550 West 120th Street, New York, NY 10027, USA \\
}
\begin{document}

\date{Accepted . Received; in original form }

\pagerange{\pageref{firstpage}--\pageref{lastpage}} \pubyear{}

\maketitle

\label{firstpage}

\begin{abstract}
The fast recurrent nova V745 Sco was observed in the 3-79 keV X-rays band with 
{\sl NuSTAR} 10 days after the optical discovery. The measured X-ray emission is consistent 
 with a collisionally ionized optically thin plasma at 
 temperature of about 2.7 keV.  
  A prominent iron line observed at 6.7 keV does not require 
 enhanced iron in the ejecta. We attribute the X-ray
 flux to shocked circumstellar material. No X-ray emission was observed at energies
 above 20 keV, and the flux in the 3-20 keV range was about 
1.6 $\times$ 10$^{-11}$ erg cm$^{-2}$ s$^{-1}$. The emission
 measure indicates an average
electron density of order of 10$^7$ cm$^{-3}$. 
%  4.7 $\times 10^6 cm^{-3}<$n$_{\rm e} >7.4 \times 10^7 cm^{-3}$.  
 The X-ray flux in the 0.3-10 keV band almost simultaneously
 measured with {\sl Swift} was about 40 times larger, mainly due to the luminous central supersoft
 source emitting at energy below
 1 keV.  The fact that the {\sl NuSTAR} spectrum 
cannot be fitted with a power law, and
 the lack of hard X-ray emission, allow us to rule out
Comptonized gamma rays, and to place
 an upper limit of the order of 10$^{-11}$ erg  cm$^{-2}$ s$^{-1}$
 on the gamma-ray flux of the nova on the tenth day of the outburst.
\end{abstract}


\section{Introduction}
 Classical novae in outburst are very luminous X-ray sources for two reasons. 
 One source of X-ray emission is the very hot white dwarf, burning hydrogen in a shell
 which is
 covered by only a thin atmosphere. In most novae the ejecta become transparent
 to X-rays before hydrogen burning has been completely quenched, so 
 the X-ray emission is extremely soft,  peaking below 1 keV,
 and very luminous, usually exceeding 10$^{37}$ erg s$^{-1}$ and often reaching
 a few times 10$^{38}$ erg s$^{-1}$ \citep[see][and numerous references
 therein]{Orio2012}. In this work we deal with the second important source of
 X-ray emission, the nova ejecta. X-ray emission due to photoionization by the
 extremely hot central source cannot be completely ruled out, but most 
 nova shells appear
  to contain collisionally ionized plasma, as expected from internal
  shocks \citep[e.g.][]{Metzger14}, even at late post-outburst
 phases \citep[e.g.][]{Rohrbach2009,
 Orio2013, Tofflemire2013}, or in symbiotic novae because the nova
 wind impacts a previous red giant wind. This
 second phenomenon occurs in the early
 phases of the outburst \citep[a typical case is RS Oph, see][]{Nelson2008}.
 The energy range of X-ray emission 
 observed from the ejected nebula varies by orders of
 magnitude in different novae. The hard X-ray emission peaked in the third week after the eruption for
 V382 Vel \citep{MukaiIshida2001}, after a week for RS Oph \citep{Sokoloski2006},
 and after only 5 days for V838 Her \citep{Lloyd1992}.

In the last four years, 
 GeV gamma rays have been detected in a handful of very luminous novae
 leaving open the possibility 
 that all other novae may be gamma ray sources at a level 
 below the luminosity detection threshold of Fermi (which is of the order
 of 10$^{36}$ erg s$^{-1}$ at a few kpc
 distance). For the symbiotic nova V407 Cyg, the gamma rays 
 were attributed to pion decay, resulting from proton-proton
 interactions in the thick red giant wind in which particles are
 accelerated by of the nova outflows \citep{Abdo2010}, 
 although \citet{Martin2013} favored instead a ``leptonic origin'', like
 in the other novae detected with Fermi (N Sco 2012, N Mon 2012, V339 Del,
 and V1369 Cen). In all these cases the gamma ray flux 
 was attributed to a combination
 of inverse Compton scattering with low
 energy photons, and bremsstrahlung, when
 the electrons are accelerated in strong shocks  in the circumstellar medium
(see Ackerman et al. 2014 and references therein).
%\citep[see][and references therein]{fermilat14}.
      A question that naturally arises is whether the hard X-ray emission of novae
 in outburst originates
 in the same medium, and from the same mechanisms, as the gamma ray emission. 
 The X-rays may be due to Compton downscattering of gamma rays \citep[see for instance][]{Masti1992}.
The X-ray flux would then increase with energy, so it is important
 to monitor the hard X-ray window.  The hard X-rays may also be due to 
a later phase in the evolution and cooling of the same shocked circumstellar plasma
 where the gamma rays originated. Hard  X-rays were
 already detected from V745 Sco with {\sl Swift} almost at the same time as the
 gamma ray emission  \citep[][]{Page14a}, but the range above
 10 keV was not observed, and this nova evolves very rapidly.

  Recent observations at radio wavelengths indicate that the mass outflow from  
 novae is not a smooth phenomenon, and that new episodes of mass ejection may
 occur long after maximum \citep[e.g.][]{Nelson2014}. In this scenario,
 a faster nova wind at some point may collide into
 a slower, initial nova wind, giving rise
 to a new burst of X-ray emission. Hard X-rays may also be associated with bipolar outflows 
 \citep[e.g. RS Oph,][]{Soko2008, Nelson2008}. Finally, it
 has been speculated that hard X-rays 
 offer diagnostics of the origin of mass outflows. They would be  
associated with outflow from the white dwarf 
 in an early outburst phase, as opposed to 
 large mass loss occurring later from the secondary  \citep{Williams2013}.
  \citet{Williams2013} noted that in the X-ray spectra of V382 Vel \citet{Orio2001},
 and later \citet{Ness2005}, did not detect iron lines,
 and suggested that the X-ray flux is emitted in the same ejecta
 as the optical ``He/N spectrum'', while the ``Fe II'' emission
 region is instead due to an outflow from the
 secondary, occurring without generating any X-ray flux. 
 If the X-ray emitting gas is from the WD,
 iron lines should not be  detectable.
 However, we do know that at least RS Oph had a clear
 iron emission feature, at least
 immediately after the outburst \citep[][]{Soko2006, Bode2006}. 
 
Because of all these reasons, it is crucial to monitor X-rays from novae since 
 the beginning of the eruption, in order to understand when the 
hard X-ray emission starts and 
 whether there is a single emission region for the gamma and 
 hard X-rays.
 The {\sl NuSTAR} satellite  is ideal for following the early X-ray emission
 of a nova shell, because of its broad energy range up to 79 keV and its
 excellent sensitivity in the iron lines region around 7 keV.
\begin{figure}
\centering
\includegraphics[width=160pt,angle=-90]{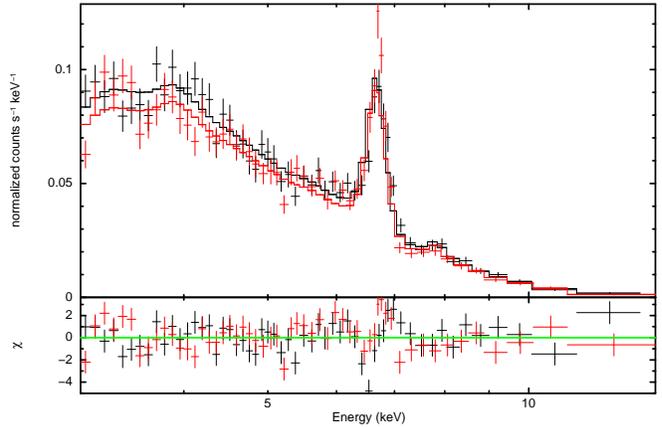}
\caption{The X-ray spectrum of V745 Sco obtained with the NuSTAR FPMA (black)
 and FPMB (red) modules. The
 3-10 keV range where most of the flux is emitted is shown in this plot.
 The fit shown is with the parameters of the first model in Table 2.}
\label{fig:Nustar}
\end{figure}
\section{A remarkable recurrent nova}
 V745 Sco is a remarkably fast nova, and one of only a handful of
 recurrent novae (RN) with  giant secondaries
\citep{Duerbeck1989, Sekiguchi1990, Williams1991}.
An orbital period of 510$\pm$20 days has been
  proposed (Schaefer 2009), but it has not
 been confirmed by OGLE data \citep{Mroz2014}. Such novae are thus also symbiotic
 stars, a spectroscopic definition of an interacting binary with a hot component and
 a cold, luminous one (a red giant, supergiant or AGB star). Symbiotics usually contain
 an accreting compact object, a white dwarf in the vast majority of cases.

 V745 Sco is the fastest Galactic nova. 
Three outbursts were observed in 1937, 1989 and 2014,
 all characterized by a very rapid decline of the light curve 
 (t$_2$=2 day and t$_3$=4 days were the times to decay by 2 and 3 magnitudes,
 respectively, in the AAVSO light curves of 2014). Each time the ejecta
 expanded at very high velocity 
\citep[][]{Duerbeck1989, banerjee14}.
 Because the outbursts evolved so rapidly, we cannot rule
 out that they occur more frequently and have not  been observed each time.
 D\"urbeck (1989) described the 1989 outbursts and 
 classified the nova as  a symbiotic system because of the TiO bands, clearly
 detected already in outburst, which are typical of an M6 III spectral type. In subsequent
 observations in quiescence the secondary was classified in the
 range M6$\pm$2 III, depending on indicators in the infrared and optical spectrum
 \citep[see brief review by][]{banerjee14}.
V745 Sco was observed in X-rays in quiescence in 2010, and appeared as a faint,
 absorbed X-ray source \citep[][]{Luna14}.
 Despite poor statistics, \citet{Luna14} performed a ``tentative fit'' 
 with an absorbed optically thin thermal plasma 
 at  temperature kT $>$10 keV and N(H)$>$1.5 $\times 10^{21}$ cm$^{-2}$ and
 attributed the X-ray emission to a disk boundary layer. 

Already at the beginning of the outburst, long before the nebular phase, a 
 strong and broad [O III] line was detected in the
 optical spectrum at 5007 \AA.  This line seems to be clearly
 associated with the red giant wind and not with the nova outflow \citep{Duerbeck1989}. A [Fe II]
 line was present from the beginning, and within two weeks the nova showed a coronal spectrum
 with other strong coronal lines of high excitation potential, including [Fe VII], [Fe X], [Fe XI], [Fe XIV], 
and [Ni XII]. 
The early evolution after the discovery on 2014, February 6, (R. Stubbings, AAVSO special
 notice no. 380) has been described in a number of Astronomer's Telegrams
\citep[see e.g.][]{Anupama2014, Page14a, Rupen2014}.
 \citet{banerjee14} have described the development of the infrared spectrum:
 the ejecta initial velocity exceeded 4000 km s$^{-1}$, and there was no 
``free expansion stage'', but  only a Sedov-Taylor expansion,
indicating a violent shock at the beginning of the outburst,
in a region of high density material already present before
 the explosion. 
  
 Fermi detected the nova in gamma rays at a 2 and 3 $\sigma$ level on the
 first day after the outburst was announced \citep{Cheung2014}. 
%\citep{fermilat14}.
 Hard X-rays 
 were also observed with Swift on the same day, and could be fitted with a thermal
 spectrum with kT$\simeq$8 keV \citep{Mukai2014}.  

\begin{figure}
\centering
\includegraphics[width=160pt,angle=-90]{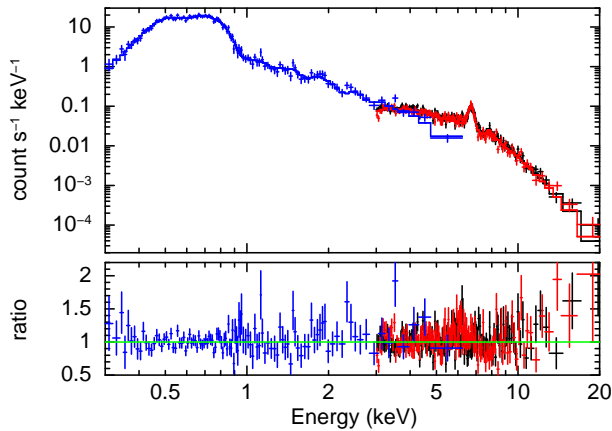}
\caption{X-ray spectra of V745 Sco obtained
with NuSTAR (3-20 keV) and with the Swift XRT (0.3-10 keV), fitted with the 
model in the second column of Table 2.}
\label{fig:wSwift}
\end{figure}

\section{The {\sl NuSTAR} Observation}
V745 Sco was
 observed with {\sl NuSTAR} 10 days after the announced optical discovery 
 on 2014 February 16 at 00:51:07 UT in a net, not continuous exposure lasting 22.5 kiloseconds,
 ending at 12:51:07 UT.
{\sl NuSTAR},  the first focusing hard X-ray telescope in 3-79 keV energy range,
 consists of two co-aligned telescopes with two focal planes, FPMA and FPMB 
%\citep{Harrison2013}.
(Harrison et al. 2013).
 The  data were processed and screened using the standard pipeline for on-axis point sources
(NUPIPELINE) in the {\sl NuSTAR} Data Analysis Software (NUSTARDAS) version 1.3.1, recently 
 included in the HEASOFT distribution. The {\sl NuSTAR} calibration database (CALDB) version 
20131223 was used throughout the analysis. The light curves and the spectra were extracted 
from the cleaned event files using NUPRODUCTS tool. 
The source photons were extracted from a circular region centered on 
the source position with a 
radius of 60 arcsec. 
A detailed background model for the source position in each of the two 
{\sl NuSTAR} telescopes was derived
 by using the nuskybgd tool (Wik et al. 2012) 
%\citep{Wik2014} 
to fit blank sky regions covering the 
entire field of view for each focal plane. This is more accurate than the usual method of 
simply scaling from the background of a nearby blank sky region, especially for faint sources, 
because it correctly accounts for gradients across the detectors.

By the time of the {\sl NuSTAR} observation, the central  supersoft X-ray 
source was the dominant source of X-ray flux (see e.g. Page et al. 2014b), but
 it is not observable in {\sl NuSTAR}'s bandpass.
The count rates
 measured with the two telescopes in different ranges are reported in Table 1.
 A luminous 
source was detected up to about 20 keV, and  above this energy 
 the observation is dominated by the background. The flux was 
1.68$\pm 0.10 \times 10^{-11}$ erg cm$^{-2}$ s$^{-1}$ 
in the 3-20 keV range. The upper limit for the flux in the 20-70 keV range is
  $\simeq 10^{-13}$ erg cm$^{-2}$ s$^{-1}$.
\begin{table}
\begin{center}
  \caption{Count rate measured with the fpma and fpmb telescopes in the
 whole NuSTAR range, and in the ``soft'' sub-ranges 3-10 keV, 3-10 keV,
 10-20 keV.} 
 \begin{tabular}{|lccc|}
\hline
 Range (keV) &  FPMA  cts/s               &    FPMB cts/s              \\
\hline
 3-79        &  0.3354$\pm$0.0040   &  0.3158$\pm$0.0039 \\
 3-20        &  0.3349$\pm$0.0039   &  0.3154$\pm$0.0038 \\
 3-10        &  0.3241$\pm$0.0038   &  0.3046$\pm$0.0037 \\
 10-20       &  0.0095$\pm$0.0007   &  0.0098$\pm$0.0007 \\
\hline
  \end{tabular}
 \end{center}
\end{table}

 We clearly detect the unresolved iron He-like triplet at 6.73$\pm$0.02 keV,
with an equivalent width of about 1 keV.
 We show in
 the next Section that the flux and plasma temperature are consistent with 
those derived analysing a Swift observation done while the {\sl NuSTAR} exposures 
were ongoing (observation id. 00033136033), however, the 
 {\sl Swift} spectrum in the SSS phase requires an extremely 
 hot and luminous stellar atmosphere, which cannot be observed at all
 in the {\sl NuSTAR} range.
\begin{table*}
\begin{center}
  \caption{Fitting models and parameters for the {\sl NuSTAR} and {\sl Swift} observations
 separately and together. The columns with the asterisk indicate
 the addition of the two oxygen emission features. 
 All the fluxes (absorbed in the 0.3-1 keV, specific flux of each
 XSPEC VAPEC model component of plasma in 
 collisional ionization equilibrium, c1 and c2,
 absorbed flux in the whole range used for the fit, and unabsorbed total 
 flux), are in units of erg cm$^{-2}$ s$^{-1}$. Component c1 is the one mainly observed
 with {\sl NuSTAR}. }
  \begin{tabular}{|lcccccc|}
  \hline
Parameter & {\sl NuSTAR}  & {\sl Swift} & {\sl Swift*}  
 & {\sl Swift} \& {\sl NuSTAR} & {\sl Swift} \& {\sl NuSTAR} *  \\ 
          & 3-10 keV      & 0.3-10 keV  &   0.3-10 keV & 0.3-79 keV & 0.3-79 keV \\
 \hline
 N(H) (10$^{21}$ cm$^{-2}$)  & 1.15$^{+0.94}_{-0.12}$    & 8.7$^{+1.2}_{-0.3}$       & 6.8$^{+1.3}_{-1.1}$ & 8.0$\pm0.4$  & 6.9$\pm0.9$ \\
 kT$_{\rm c1}$  (keV) & 2.66$^{+0.09}_{-0.14}$  & 1.62$^{+0.14}_{-0.11}$    & 2.77$^{+0.87}_{-0.50}$ & 2.76$^{+0.07}_{-0.10}$ & 2.76$^{+0.08}_{-0.10}$ \\
 kT$_{\rm c2}$ (keV)  &       & 0.068$^{+0.001}_{-0.013}$ & 0.86$^{+0.12}_{-0.16}$ & 0.83$^{+0.11}_{-0.12}$ & 0.87$^{+0.12}_{-0.13}$  \\
 Fe/Fe$_\odot$   & 0.51   &  0.51    &   0.51      & 0.51  &  0.51 \\
 T$_{\rm eff}$ (10$^5$ K) &          & 9.81$^{+0.04}_{-0.06}$    &  10.0$\pm$0.2 & 9.79$\pm$0.04    & 10.0$^{+0.2}_{-0.1}$ &  \\
 F(0.3-1 keV)      &           &  1.71 $\times 10^{-10}$ & 1.81 $\times 10^{-10}$  & 1.66 $\times 10^{-10}$ & 1.81 $\times 10^{-10}$ \\
 F$_{\rm c1}$      & 1.60$\times 10^{-11}$  & 4.02$\times 10^{-11}$ & 3.95$\times 10^{-11}$   & 2.97$\times 10^{-11}$ & 3.04 $\times 10^{-11}$ \\
  F$_{\rm c2}$     &           & 4.31$\times 10^{-11}$ & 1.70$\times 10^{-11}$  & 1.91 $\times 10^{-11}$ & 1.90 $\times 10^{-11}$ \\
 F$_{\rm tot}$     & 1.60$\times 10^{-11}$  & 2.11 $\times 10^{-10}$  & 2.26$\times 10^{-10}$ & 2.16 $\times 10^{-10}$ & 2.39 $\times 10^{-10}$ \\
 F$_{\rm tot,u}$   & 1.62$\times 10^{-11}$  & 3.66 $\times 10^{-8}$  & 3.66 $\times 10^{-8}$  & 7.39 $\times 10^{-8}$ &  3.81 $\times 10^{-8}$ \\
 $\chi ^2$/d.o.f.  & 1.3       &    2.3   & 1.06      & 1.6      & 1.3 \\
\hline
  \end{tabular}
 \end{center}

\end{table*}

\section{Spectral fits and additional Swift data}
 A fit to the NuSTAR spectrum 
 is shown in Fig. 1 and Table 2, using an XSPEC ``VAPEC'' model of
plasma in collisional ionization equilibrium at kT=2.66 keV,  flux
 1.6 $\times 10^{-11}$ erg cm$^{-2}$
s$^{-2}$ and iron abundance [Fe/Fe$_\odot$]=0.51. This best fit
 yields a reduced $\chi^{2}$ value of 1.3. 
 We were not able to obtain a better fit with different models, even  
by adding more model components. 
 From the emission measure fit we can obtain the
 average electron density of the ejecta emitting the X-rays,
 by making an assumption about the distance traveled by the
 ejecta. The work  of Banjeree et al. (2014) is
 very useful, because these authors measured the velocity on 
 different dates. They comment that the velocity law
 can be fitted with a third degree polynomial law, but 
they do not specify how their fit in their figure 3 was
 obtained. To find the range
 of interest of the electron density we  can simply assume that radius of
 an idealized spherical shell would be smaller that 
 reached with the initial epansion velocity
 (4825 km s$^{-1}$), but larger than reached with the velocity at day
10.3 (the time of the {\sl NuSTAR} observation,
 that is 1930 km s$^{-1}$. We thus
 obtain a value of  n$_{\rm e}$ between 4.7 $\times
 10^6$ cm$^{-3}$ and 7.4 $\times
 10^7$ cm$^{-3}$.  This range is only indicative
 of the average value, and is obtained with the assumption that
the flux was emitted in a spherical shell around the WD, of
 homogeneous density. 
 The corresponding mass of the ejecta, if the hard X-ray
 emission region represents all the ejected mass,
 would be 5 $\times 10^{-8}$ M$_\odot$.
We note that a power law component does not improve the fit, and clearly the data cannot
 be well fitted with a power law. The prominence of the iron feature is only due to 
the sensitivity of {\sl NuSTAR} in the range around 7 keV, but the iron abundance 
 we obtained from the best fit is only [Fe/Fe$_\odot$=0.51. We kept this abundance
 value fixed in
 the other fits described below. 

 {\sl Swift} data were taken at semi-regular intervals of
 half a day or less, and those obtained just a few hours before
 the {\sl NuSTAR} exposure (2014 February 16 1:35:17 UT for an exposure time
 of 1038 s,  observation 00033136033) explain the difficulty
 in finding an optimal fit: the central source is more luminous than the nova shell and 
the lower limit
 of the {\sl NuSTAR} bandpass is higher than the energy at which
 most of the X-rays are emitted. 
The soft part of the spectrum is very luminous and probably due to 
 more than one emission mechanism, so 
 the range around 3 keV is not well constrained.
 The full {\sl Swift} dataset describing the nova evolution with snapshot exposures taken even three times
 a day is being analysed by the {\sl Swift} team with several coauthors (Page et al. 2014c,
 in preparation), but here suffice it to say that we checked two additional exposures taken a few
 hours after the {\sl NuSTAR} one and did not find significant spectral evolution in
 this short time. 
The large luminosity of the supersoft source
 (see Fig. 2) is still below a critical threshold of about 
 50 cts s$^{-1}$ that may cause pile-up for the supersoft
 sources in {\sl windowed timing mode}, used for this observation.
  In Table 2 we report the results of two fits of the {\sl Swift}
 spectrum of this observation. In the first one we used
 a stellar atmosphere at 930,000 K (Rauch 2010, model ), a low temperature plasma
 VAPEC model component
 in XSPEC at 68 eV, and a third a component, VAPEC  at about 1.6 keV, obtaining 
 a value 2.3 for the reduced $\chi ^{2}$.
 We cannot improve the fit with
 additional components, but it improves
by using a blackbody with an oxygen absorption edge at 0.87 keV
  of arbitrary depth. 
 An even better fit was obtaining by superimposing additional emission features 
 of oxygen on the VAPEC models,  H-like Lyman $\alpha$ at 0.65 keV and a blended
 He-like triplet at 0.57 keV. The fit results statistically acceptable, with $\chi^2=1.1$.
  The total flux is 2.26 $\times 10^{-10}$ erg cm$^{-2}$ s$^{-1}$, largely due to the supersoft
 source, with an unabsorbed flux 3.66 $\times 10^{-8}$ erg cm$^{-2}$ s$^{-1}$.   
 The physical reason behind the need to add these ``artificial'' features is 
 not due to an unusually high oxygen abundance in the plasma component. 
 We did not have available
 grids of atmospheric models with a lower oxygen abundance, so that the
 oxygen absorption features were too deep and needed to be ``corrected''. Since
 atmospheric models with appropriate abundances are not available yet,
 we resorted to this arbitrary addition.
 
 The best fit to the {\sl Swift} and the {\sl NuSTAR}
 spectra together is shown in Fig. 2, with  
 the parameters in Table 2. We can constrain
 the hotter component and the value of N(H) using both sets of X-ray telescopes,
 and obtain N(H)=(6.9$\pm$0.9) $\times$ 10$^{21}$ cm$^{-2}$, while the hot plasma has kT=2.76 keV.
 We note that N(H) is higher than the value $\simeq 4.8 \times 10^{21}$ cm$^{-2}$
 that one would expect applying e.g. the relationship 
 derived by \citet{Feryal} to the interstellar reddening E(B-V)=0.7 given
 in Banerjee et al. (2014), impling a certain amount of intrinsic absorption,
 likely due to the red giant wind.

 \section{Conclusions}
 The observation of the very fast RN V745 Sco with {\sl NuSTAR} 10 days after the discovery
 of the optical outburst and the observed maximum indicates that the hard X-ray emitting
 region at this stage had already cooled. We rule out plasma components hotter than
 2.8 keV. Hard X-ray emission due to Comptonized gamma rays can also
 be ruled out because it would produce a power law spectrum
 with copious flux above 20 keV \citep[e.g.][]{Masti1992}, which was not observed
 in the high energy band of {\sl NuSTAR}. 

 The threshold detection flux of Fermi is of the order of 10$^{-10}$ erg cm$^{-2}$ s$^{-1}$, so
 we do not know whether there was gamma-ray emission with flux below this level, but
 hard X-rays are also an important indicator of gamma rays
 because the flux due to Comptonized down-scattered photons should be measurable.
 The optical depth of Compton scattering is $$t_{\rm e}=n_{\rm e} \sigma_{\tau} {\rm D},$$
 where n$_{\rm e}$ is the electron density and the
 Thomson cross section is $\sigma_{\tau}$=6.6524 $\times 10^{-25}$ cm$^2$. Adopting n$_{\rm e} \simeq 
 \times 10^7$ cm$^{-3}$ derived above, and assuming that the photons have to cross a region
 of depth D as large as the distance to which the ejecta at have
expanded in 10 days. Given the
 range of velocities from the first and tenth day (Banerjee et al. 2014,
 see above), 1.7 $\times 10^{14}$ cm$<$D$< 4.2 \times 10 ^{14}$ cm, so 
 we derive t$_{\rm e}$ to be in the range
 1.1-2.8 $\times 10^{-3}$. Optical depth t$_{\rm e} < 1$ indicates  
 that the ejecta are optically thin to the Comptonized radiation,
 which should be clearly detectable for V745 Sco. 

Do we have enough material in the shell to degrade
 the gamma-ray photons to X-rays with a significant Comptonized flux?
 \citet{Masti1992} calculated the X-ray light curve due to Comptonized
 gamma rays in the case of the radioactive decay of $^{22}$Na, at
 the specific fixed energy of the line at 1.22 MeV, which is two
 orders
 of magnitude lower than the lower range of the  
continuum gamma-ray emission of V745 Sco detected with Fermi. Despite the different
 gamma-ray range,  
these authors'   Monte Carlo simulations should be relevant also for V745 Sco. 
 In \citet{Masti1992} a peak of X-ray
emission of order of one hundredth the value of the
 gamma-ray flux was predicted 200 days
 after the burst of gamma-ray emission, for ejecta with mass of 10$^{-5}$ M$_\odot$ and
 a residual velocity of 10 km s$^{-1}$. The time
 to reach peak emission is directly proportional to the ejected mass
 and inversely proportional to its expansion velocity. Since the ejecta mass in
 a fast RN like V745 Sco is likely to be at least
 a factor of 10 lower than above, and the ejecta velocity  after
 10 days was still 200 times
 higher, a straightforward analogy would imply
 that the maximum emission of Comptonized
 X-rays for V745 Sco is expected 
 within a few hours. The flux in our case should decay very quickly, like in the case
 analysed by \citet{Masti1992}, 
 peaking at a value around one hundredth the value of the gamma-ray flux. 
 The upper limit for X-ray flux above 50 keV measured
 with {\sl NuSTAR} is about 10$^{-13}$ erg cm$^{-2}$ s$^{-1}$, roughly translating into  
 an upper limit estimate of the gamma ray  
 flux 10 days after the outburst of V745 Sco of the order of 10$^{-11}$ erg cm$^{-2}$ s$^{-1}$. 

   We attribute the X-ray flux we measured with {\sl NuSTAR} to shocks in the
 ejected shell, most likely as they impacted the red giant wind of the 
  companion, like in the cases of V407 Cyg and RS Oph. 
 We cannot tell at
 this stage whether the emission was due to the same region 
where the gamma rays originated on the first day, 
 but it appears that in RN nova
 shells there are multiple regions of shocked material at different temperatures.  

\section*{Acknowledgements}
We wish to acknowledge the support of the following awards and funding agecies:
M. Orio of ASI (Italian Space Agency) through
 an INAF-ASI grant,
 J.L. Sokoloski of the NSF award AST-1211778,
 and K.L. Page of the UK Space Agency funding  
the UK Swift Science Data Centre.

\bibliographystyle{mn2e}
\bibliography{biblio}
%Harrison, F. A., Craig, W W., Christensen, F. E., et al. 2013, ApJ, 770, 103
%Wik, D. R., Hornstrup, A., Molendi, S., et al. 2014, ApJ, submitted (arXiv:1403.2722)

%\label{lastpage}

\end{document}